\documentclass[aps,prl,showpacs,twocolumn,superscriptaddress]{revtex4}
\usepackage{times,xspace}
\usepackage{amsbsy,amssymb,amsmath,bm}
\usepackage{graphicx,color,epsfig,rotate}
\usepackage{amsfonts}
\begin{document}
\title{Electromagnetic radiation from vortex flow in type-II superconductors}
\author{L.N. Bulaevskii}
\affiliation{Los Alamos National Laboratory, Los Alamos, New Mexico 87545}
\author{E.M. Chudnovsky }
\affiliation{Physics Department, CUNY Lehman
College, 250 Bedford Park Boulevard West, Bronx, New York
10468-1589
 }

%\thanks{}
% \altaffiliation{}
\date{\today}

\begin{abstract}
We show that a moving vortex lattice, as it comes to a crystal
edge, radiates into a free space the harmonics of the washboard
frequency, $\omega_0=2\pi v/a$, up to a superconducting gap,
$\Delta/\hbar$. Here $v$ is the velocity of the vortex lattice and
$a$ is the intervortex spacing. We compute radiation power and
show that this effect can be used for generation of terahertz
radiation and for characterization of moving vortex lattices.
\end{abstract}
%\keywords{}
\maketitle

Effects at the washboard frequency in the flux-flow regime of type
II superconductors were first reported by Fiory \cite{Fiory}.
Imposing an rf current at the harmonics of $\omega_{0}=2\pi v/a$
on top of the dc transport current, Fiory observed steps in the
I-V characteristics of superconducting aluminum films. Here $v$ is
the vortex velocity and $a$ is the vortex lattice parameter in the
direction of ${\bf v}$. Similar observations were made in
high-temperature superconductors \cite{Harris,Togawa}. Larkin and
Ovchinnikov \cite{LO}, and Schmid and Hauger \cite{SH} have
demonstrated that the effect originates from  the coherent action
of defects as they are passed periodically at the washboard
frequency $\omega_0$ by moving vortices. It is analogous to
Shapiro steps in Josephson junctions in the presence of the ac
current \cite{Kulik}. By analogy with the radiation from a
Josephson junction \cite{Joseph,Dm,Lan}, the inverse effect of the
electromagnetic radiation from a moving vortex lattice at the
washboard frequency should be expected. Radiation produced by a
single vortex or vortex bundle crossing the edge of a
superconductor was discussed by Dolgov and Schopohl \cite{DS}.
They used the transition radiation approach to calculate the
energy and spectrum of the broad-band electromagnetic pulse
emitted by the bundle. In this Letter we show that the situation
for a moving vortex lattice is completely different. As vortices
come to the surface periodically with the frequency $v/a$, they
radiate into free space the harmonics of $\omega_0$ up to a
frequency corresponding to the superconducting gap,
$\Delta/\hbar$. Unlike the rf component in the transport current
produced by defects \cite{Fiory,Harris,Togawa,LO,SH}, the
electromagnetic radiation from the surface does not disappear but
becomes stronger when disorder weakens. It is generated by
oscillating electric and magnetic fields of vortices near the
surface and propagates into free space due to the continuity of
tangential components of the fields at the surface. We study cases
of ideal and disordered vortex lattice in small and large crystals
as compared to the radiation wavelength. Radiation power is
derived and shown to be within experimental reach. For a large
crystal the effect is proportional to the radiating surface, while
for a small crystal it is quadratic on the surface, that is,
superradiant. The challenge for experiment is to satisfy the
requirements of sufficient speed and appreciable correlation
length of translational order. If these conditions are satisfied,
the effect can be used for generation of electromagnetic radiation
well into the terahertz range.

We study electromagnetic radiation from the boundaries, $x = 0,
-L_x$, of a type-II isotropic superconductor, located at $-L_x < x
< 0$, into a free space at $x >0$ and $x < -L_x$, see Fig. 1. The
applied magnetic field, ${\bf H}_0$, along the $z$ axis is assumed
to satisfy $H_{c1}\ll H_0\ll H_{c2}$, with $H_{c1}$ and $H_{c2}$
being the first and the second critical fields. The transport
current along the $y$-axis results in the motion of the vortex
lattice at a speed ${\bf v}$ along the $x$-axis. We consider large
${\bf v}$ dominated by the Lorentz force and vortex drag, so that
the effect of the surface Meissner current and the back effect of
the radiation on the lattice motion can be ignored.
\begin{figure}[ptb]
\begin{center}
\includegraphics[width=0.32\textwidth,clip]{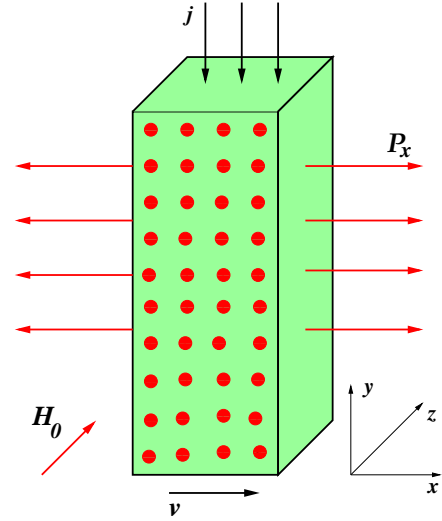}
\end{center}
\par
\vspace{-0.19in}\caption{(Color online) A moving vortex lattice
radiates electromagnetic waves into a free space. Arrows indicate
directions of the applied magnetic field ${\bf H}_0$, transport
current ${\bf j}$, vortex lattice velocity ${\bf v}$, and the
Poynting vector ${\bf P}_x$. Vortices are shown by circles.}
\label{Fig1}
\end{figure}
To obtain the radiation power we need to find electromagnetic
fields, $B_z$ and $E_y$, at the surfaces $x = 0$ and $x = -L_x$.
At $v \ll c$ the fields inside the superconductor can be obtained
from a quasistatic approximation (${\bf r}=x,y,z)$:
\begin{eqnarray}
&&{\bm \nabla} \times {\bm \nabla} \times {\bf
B}+\frac{1}{\lambda^{2}}{\bf B}= \label{B1}\\
& & \frac{\Phi_0{\bf
e}_z }{\lambda^2}\sum_{\bf n}\int dz\delta[{\bf r}-{\bf
r}_{n}(t,z)]\Theta[-x_{\bf n}(t)]\Theta[L_x + x_{\bf n}(t)]
\nonumber \\
&&c[{\bm \nabla}\times {\bf E}]=-{\partial {\bf B}}/{\partial
t}\,. \label{E1}
\end{eqnarray}
where ${\bf e}_z$ is the unit vector along the $z$-axis,
$\Theta(x)$ is a unit step function, $\lambda$ is the London
penetration length, and ${\bf r}_{\bf n}(t,z)$ is the position of
the center of the ${\bf n}=(n,p)$-th vortex line in the vortex
lattice. Consider first a square lattice with the lattice spacing
$a = \sqrt{\Phi_0/B} \ll \lambda$. We shall assume that the
transport current $j$ is sufficient to drive vortices in the
laminar (Bragg glass) regime \cite{KV,DG}:
\begin{equation}
x_n(t) = an + vt\,, \ \ y_p(z) =ap+\delta_p(z)\,.
\label{xy}\end{equation} Here $n,p$ are integers and $\delta_p(z)$
accounts for disorder in the position of vortices along the
$y$-axis. Strictly speaking, vortices do not form an ideal
periodic lattice along the $x$-axis. Rather, in the absence of
topological defects, moving vortices adjust to quenched disorder
by periodically replacing each other at some favorable points,
$x_n(0,p,z)$, with a period $T=a/v$ \cite{DG},
$$x_n(t,p,z)=x_n(0,p,z)+[x_{n+1}(0,p,z)-x_n(0,p,z)]f(t).$$
Here $f(t)$ is a periodic function with the period $T$, satisfying
$f(t) \approx t/T$ inside the interval $[0,T]$. One can show,
however, that at $a\ll\lambda\,$ Eq.~(\ref{xy}) is sufficient to
describe the motion of Bragg glass along the $x$-axis -- the
magnetic field at the boundary (that determines radiation) is only
weakly affected by disorder in $x_n(0,p,z)$. Phase shifts in
otherwise perfectly periodic process are introduced only by
time-dependent thermal fluctuations. This is similar to Josephson
oscillations, see Ref.~\onlinecite{IKulik}. The similarity of the
vortex-induced radiation to Josephson oscillations also follows
from the fact that the washboard frequency satisfies the relation
$\hbar\omega=2eV$, where $V$ is the voltage drop for one period of
the vortex lattice in the direction of the current. This is a
consequence of the expression for the dc electric field in that
direction, $E_y^{(dc)}=(v/c)B$. Note that delta-functions in the
right hand side of Eq.~(\ref{B1}) correspond to point-like vortex
cores. In reality the cores have radius $\xi$, the superconducting
correlation length, which will be accounted for in the following.
Here we will ignore thermal fluctuations that lead to the
broadening of radiation lines.

For $L_x\gg\lambda$ one can prove that the two surfaces, $x = 0$
and at $x = -L_x$, can be studied independently. We, therefore,
begin with considering a semi-infinite superconductor at $x<0$.
The fields inside the superconductor ($x <0$), must be matched by
solutions of Maxwell equations in the free space, ($x > 0$). The
total fields inside the superconductor are
\begin{eqnarray}\label{total}
B_z(t,{\bf r}) & = & B_{zv}(t,{\bf r})+B_{z0}(t,{\bf r}), \nonumber \\
E_y(t,{\bf r}) & = & E_{yv}(t,{\bf r})+E_{y0}(t,{\bf r})\,,
\end{eqnarray}
where $B_{zv}$ and  $E_{yv}$ are magnetic and electric fields
produced by vortices, while $B_{z0}$ and  $E_{y0}$ are solutions
of Eqs.\ (\ref{B1}) and (\ref{E1}) with a zero right-hand side in
Eq.~(\ref{B1}). For $B_{zv}$ one obtains
\begin{eqnarray}
& & B_{zv}(\omega,{\bf k}) = \sum_{\bf n}\int
dze^{ik_zz}\frac{\Phi_0}{1+\lambda_{\omega}^2{\bf k}^2}\;
\frac{e^{-i\omega an/v-ik_yy_p(z)} }{i\omega-ik_xv-\epsilon}
\nonumber \\
& &= \sum_{m,p}\int dz\,e^{ik_zz}\frac{2 \pi
v\Phi_0}{a(1+\lambda_{\omega}^2{\bf k}^2)}\;
\frac{e^{-ik_yy_p(z)}\delta(\omega -
m\omega_0)}{i\omega-ik_xv-\epsilon}\,, \nonumber \\ \label{bzv}
\end{eqnarray}
where ${\bf k}=(k_x,k_y,k_z)$. The summation over $n$ was carried
out with the help of the relation
\begin{equation}
\sum_{n}\exp[-(i\omega a/v)n]=2\pi\sum_{m=1}^{M}\delta(\omega
a/v-2\pi m),
\end{equation}
where the upper limit $M\sim a/\xi$ is due to the nonzero size,
$\xi$, of the vortex core. The frequency dependent London length
is given by $\lambda_{\omega}^{-2}=\lambda^{-2}-k_{\omega}^2+4\pi
k_{\omega}\sigma_q/c$, where $\sigma_q$ is the quasiparticle
conductivity and $k_{\omega}=\omega/c$. Integrating
Eq.~(\ref{bzv}) over $k_x$ and using inequality $a\ll\lambda$ we
obtain the amplitude of the oscillating magnetic field at the
boundary $x=0$ at $\omega=\omega_m=m\omega_0$:
\begin{equation}
B_{zv}(\omega,x=0,k_y,k_z)=\sum_{p}\int
dz\frac{\Phi_0ve^{i[k_zz-k_yy_p(z)]}}{2i\omega\lambda_{{\omega}_m}a}\,.
\label{bzv1}
\end{equation}
For the electric field of the vortex
lattice, with the help of Eq.~(\ref{E1}) and ${\bm
\nabla}\cdot{\bf E} = 0$ one obtains $E_{yv}(\omega,{\bf k} )=
(vk_x^2/(c{\bf k}^2)B_{zv}(\omega,{\bf k})$. Solutions of the
homogeneous equations are
$E_{y0}(\omega,x,y)=-ik_{\omega}\lambda_{\omega}^2B_{z0}(\omega,x,y)$ and
\begin{equation}
B_{z0}(\omega,x,y)=\int
\frac{dk_y}{2\pi}A(\omega,k_y)e^{(\lambda_{\omega}^{-2}+k_y^2)^{1/2}x+ik_yy},
\end{equation}
where $A(\omega,k_y)$ is determined by the continuity of the
tangential components of the fields at the boundaries of the
superconductor.

Maxwell equations determine the relation between $B_z$ and $E_y$
everywhere in free space $x>0$. To find this relation at the
boundary, $x\rightarrow +0$, we follow the derivation outlined in
Ref.~\onlinecite{bk}. We assume that at $x > 0$ there is {\it only
outgoing} electromagnetic wave. Then the electric field
$E_y(\omega,{\bf r})=\int dt\exp(i\omega t)E_y(t,{\bf r})$ in free
space is completely determined by its value at the boundary
through [${\bf k}_{\perp}=(k_y,k_z)$]
\begin{eqnarray}
&&E_y(\omega,{\bf r} )=\int\frac{d{\bf k}_{\perp}
}{(2\pi)^2}E_{y}(\omega,x=0,{\bf k}_{\perp})e^{i{\bf
k}_{\perp}{\bf r}}
\nonumber \\
&&\times\left\{\exp[i(k_{\omega}^2-{\bf k}_{\perp}^2)^{1/2}{\rm
sign}(\omega)x] \Theta(k_{\omega}^2-{\bf k}_{\perp}^2)\right.
\nonumber \\
&& \left.+\exp[-({\bf k}_{\perp}^2-k_{\omega}^2)^{1/2}x)]
\Theta({\bf k}_{\perp}^2-k_{\omega}^2)\right\} \,. \label{ey}
\end{eqnarray}
The first term in the right hand side is due to radiated waves,
while the second term comes from the waves decaying exponentially
away from the boundary. Using this relation, Eq.~(\ref{E1}) and
$\nabla\cdot {\bf E}=0$, we express the magnetic field in free
space via $E_y(\omega,x=0,y,z)$ and finally obtain the relation
between $B_z(\omega,x=0,y,z)$ and $E_y(\omega,x=0,y,z)$ at the
boundary:
\begin{eqnarray}
&&B_{z}(\omega,0,{\bf k}_{\perp})=\zeta(\omega,{\bf k}_{\perp})E_{y}(\omega,0,{\bf k}_{\perp})\,,
\label{BoutEout} \\
&&\zeta(\omega,{\bf k}_{\perp}) \equiv
\frac{|k_{\omega}|\Theta(k_{\omega}^2-{\bf
k}_{\perp}^2)}{\sqrt{k_{\omega}^{2}-{\bf k}_{\perp}^{2}}}-
\frac{ik_{\omega}\Theta({\bf
k}_{\perp}^2-k_{\omega}^2)}{\sqrt{{\bf
k}_{\perp}^{2}-k_{\omega}^{2}}} \label{z2}\,.\nonumber
\end{eqnarray}
For the left boundary, $x=-L_x$, one should reverse the sign of
$\zeta$ in Eq.~(\ref{BoutEout}).

The above relations allow one to determine the fields $B_{z0}$ and
$E_{y0}$ and, finally, the total fields $B_z$ and $E_y$ at the
boundary as well as the Poynting vector of the radiation.
Radiation power at frequency $\omega$ outside the superconductor
at $x > 0$ (to the right) is given by
\begin{eqnarray}
&&  \mathcal{P}^{{\rm r}}_{\rm rad}(\omega)=\frac{c}{4\pi}\int
dydz\operatorname{Re}[E_{y}e^{- i\omega t}]
\operatorname{Re}[B_{z}e^{-i\omega t}] \nonumber \\
&& =\frac{c}{8\pi}\int \frac{d{\bf
k}_{\perp}}{(2\pi)^2}\operatorname{Re}\left[ \zeta
^{-1}(\omega,{\bf k}_{\perp})\right] |B_z(\omega,0,{\bf
k}_{\perp})|^{2}\,, \nonumber \\ \label{Poy}
\end{eqnarray}
where integration is over $|{\bf k}_{\perp}|<|k_{\omega}|$. Using
Eqs.~(\ref{total}) and (\ref{BoutEout}) we obtain for the
amplitude of the oscillating magnetic field
\begin{equation}
E_y(\omega,0,{\bf k}_{\perp})=\frac{E_{yv}(\omega,0,{\bf
k}_{\perp})+i\lambda_{\omega}k_{\omega}B_{zv}(\omega,0,{\bf
k}_{\perp})} {1+i\lambda_{\omega}k_{\omega}\zeta(\omega,{\bf
k}_{\perp})}
\end{equation}
with $\lambda_{\omega}k_{\omega}\ll 1$ and $E_{yv}(\omega,0,{\bf
k}_{\perp})\ll \lambda_{\omega}k_{\omega}B_{zv}(\omega,0,{\bf
k}_{\perp})$ at $\omega = m\omega_0<\Delta/\hbar$. The power of
the radiation into free space to the right or to the left of the
superconductor becomes
\begin{eqnarray}
&&{\cal P}_{\rm rad}^{{\rm r,l}} \approx
\frac{ck_{\omega}}{8\pi}\int\frac{(d{\bf k}_{\perp}/4\pi^2)}
{\sqrt{k_{\omega}^2-{\bf k}_{\perp}^2}}\{|\lambda_{\omega}k_{\omega}B_{zv}(\omega,0,{\bf k}_{\perp})|^2
\nonumber \\
&&\pm 2{\rm
Im}[\lambda_{\omega}k_{\omega}B_{zv}(\omega,0,{\bf k}_{\perp})E_{yv}(\omega,0,{\bf k}_{\perp})]\}\,
.  \label{Poynting}
\end{eqnarray}
We see that the radiation power is slightly stronger to the right
(in the direction of the vortex motion), though the difference is
small. In the following we neglect this difference.

Substituting the amplitude of the magnetic field at the boundary,
$B_{zv}(\omega,0,{\bf k}_{\perp})$ of Eq.~(\ref{bzv1}), into the
expression (\ref{Poynting}) for the radiation power, and averaging
over disorder in the vortex lattice, we obtain the power,
\begin{eqnarray}
{\cal P}_{\rm rad}(\omega_m) & = &
L_yL_z\frac{\Phi_0^2v^2k_{\omega}}{32\pi ca^2}\int\frac{d{\bf
k}_{\perp}}{(2\pi)^2} \frac{S({\bf k}_{\perp})}
{\sqrt{k_{\omega}^2-{\bf k}_{\perp}^2}} \label{power-next}\\
S({\bf k}_{\perp}) & = & \sum_{p}\int
\frac{dz}{a}\exp[ik_zz-ik_yy_p(z)]\,,
\end{eqnarray}
where $S({\bf k}_{\perp})$ is the structural factor of the moving
vortex lattice. It is peaked at $k_z=0$ and $k_y=2\pi n/a$, and
has the following properties: $S(0,0)={L_yL_z}/{a^2}\,,\; \int
{d{\bf k}_{\perp}}S({\bf k}_{\perp})=(2\pi/a)^2$. If translational
correlations in the vortex lattice decay exponentially, then at
$L^{-1}_{y,z} \ll k_{y,z}\ll (al_{y,z})^{-1}$ one should use
$S(k_y,k_z)=\ell_y \ell_z$, where $\ell_y$ and $\ell_z$ are
dimensionless transversal and longitudinal correlation lengths
expressed in units of the intervortex spacing $a\;$
($\ell_z\gg\ell_y$) \cite{Blatter}.

Consider first the case of weak disorder, when correlation lengths
exceed dimensions of the crystal, $a\ell_y\gg L_y$ and $a\ell_z
\gg L_z$, and when the wavelength of the radiation is small
compared to the crystal, $c/\omega_m \ll L_y,L_z$. With account of
the above properties of the structural factor, integration over
${\bf k}_{\perp}$ in Eq.~(\ref{power-next}) gives for the
radiation power at $\omega = \omega_m$
\begin{equation}
\mathcal{P}_{\rm rad}(\omega_m) =L_yL_z  \frac{v^2B^2}{32\pi c}\,,
\label{clean}\end{equation} where we neglected insignificant
dynamical terms in $\lambda_{\omega}$. An amusing feature of the
above result is that each electromagnetic mode up to $\omega_m
\sim \Delta/\hbar$ satisfying conditions $c/\omega_m \ll L_y,L_z$
makes equal contribution, $v^2B^2/(32\pi c)$, to the radiation
power per unit square of the radiating surface. The total power,
summed up over all modes, is of order $\mathcal{P}^{(tot)}_{\rm
rad}/L_zL_y = (M/32\pi)(v^2B^2/c) \sim
[1/(16\sqrt{2\pi})](v^2/c)B^{3/2}H_{c2}^{1/2}$. For weak disorder
but small crystal (or lower frequency $\omega_m$), $ L_y,L_z \ll
c/\omega_m $, we obtain
\begin{equation}\label{SR}
\mathcal{P}_{\rm rad}(\omega_m) = \left[L_yL_z \frac{v^2B^2}{32\pi
c}\right]\frac{L_yL_z\omega_m^2}{c^2}\,.
\end{equation}
Quadratic dependence of the radiation power on $L_y$ and $L_z$ is
due to superradiance, that is, coherent radiation by vortices
positioned within the radiation wavelength.

Stronger disorder results in smaller correlations lengths,
$a\ell_y \ll L_y, (c/\omega_m)$ and $a\ell_z \ll L_z,
(c/\omega_m)$. In this case the power of the radiation is reduced:
\begin{equation}
\mathcal{P}_{\rm rad}(\omega_m) = \left[L_yL_z \frac{v^2B^2}{32\pi
c}\right]\frac{a^2\ell_y\ell_z\omega_m^2}{c^2}\,.
\end{equation}
The coherent radiation takes place within the surface area of size
$(l_ya)(l_za)$, while the total radiation power is a sum of
independent contributions from such areas. Note, that
$\ell_y,\ell_z$ are correlation lengths for the {\it moving}
vortex lattice. Due to the effect of dynamical reordering
\cite{KV,DG,Yaron,Pardo}, they must be greater than the ones for
the static lattice. Thus measurements of the electromagnetic
radiation from moving vortex lattices provide information on
translational correlation lengths and their dependence on
velocity.

Static square lattices have been observed in high-temperature
superconductors: La$_{1.83}$Sr$_{0.17}$CuO$_{4+\delta}$ in fields
above 0.4 T \cite{Gilardi} and YBa$_2$Cu$_3$O$_{7}$ in fields
above 11 T \cite{Brown}. Square vortex lattices have been also
observed in borocarbides \cite{boro}. There is little information,
however, on whether the rectangular symmetry is preserved in a
vortex lattice moving at a high speed. For a triangular vortex
lattice $x_{np}$ in Eq.~(\ref{xy}) must be replaced by
$x_{np}=an+vt+a[1+(-1)^p]/4$. Then Eq.~(\ref{bzv}) acquires an
additional factor $[1+\exp(i\pi m)]$ so that only even harmonics,
$2m\omega_0$, of the washboard frequency are present in the
radiation; their intensity being 1/4 of that for a square lattice.
For a vortex liquid the integrals must be dominated by
correlations at small distances, that is by $k_y$ of order $1/a$,
and the radiation becomes suppressed by a factor
$a^2\omega_m^2/c^2$, i.e., it is practically absent.

Energy burst from a bundle of $\Lambda = L_xL_y/a^2$ vortices
crossing the edge of a superconductor has been discussed by Dolgov
and Schopol \cite{DS}. They correctly noticed that the radiation
spectrum from the vortex lattice consists of discrete lines and
can be superradiant. According to Ref. \onlinecite{DS}, the
transient radiation of a single bundle, as in the case of a single
vortex, peaks at $(a/\sqrt{3}\lambda)\omega_{0}$, which is
approximately the inverse time needed for the vortex field of
diameter $\lambda$ to cross the crystal edge. We found completely
different spectral power of the electromagnetic radiation from a
moving vortex lattice.

To obtain significant radiation power one needs high vortex
lattice velocity $v$. This velocity is limited by dissipation
(heating). Pulse technique helps to diminish heating and reach
higher velocities \cite{Gilardi,Brown} up to the critical velocity
$v^*$ arising from Larkin-Ovchinnikov instability \cite{LO2}. The
latter occurs due to the decrease of the vortex drag coefficient
with $v$ as $\eta(v)=\eta(0)({1+v^2/{v^*}^2})^{-1}$. Such a
behavior of $\eta(v)$ is caused by the escape of quasiparticles
from normal cores due to the effect of the electric field. It
leads to a negative slope in the I-V characteristics at velocities
above $v^*$. The latter depends on the inelastic scattering of
quasiparticles and decreases on cooling. Experimental study
\cite{D1} of this instability in Nd$_{1.85}$Ce$_{0.15}$CuO$_x$
gives $v^*=8$ m/s at 8 K in the field 120 mT. At such velocity we
estimate the radiation power as 2 $\mu$W/cm$^2$ at $\omega \sim
\Delta/\hbar$ in the field of 1 T for a square vortex lattice in
the case of weak disorder, Eq.~(\ref{clean}). Much higher value of
the critical velocity, $v^*=1.2\cdot 10^3$ m/s, was obtained in
YBa$_2$Cu$_3$O$_{7-\delta}$ films at 72 K in the field of 1.8 T
\cite{D2}. At such speed and field the radiation power from a
weakly disordered square lattice would be of order 0.1 W/cm$^2$.

To obtain the efficiency of the radiation source, we write the
dissipation power as ${\cal P}_{{\rm
dis}}=({BL_xL_y}/{\Phi_0})(\eta v^2L_z/2)$, where $B =\Phi_0/a^2$,
$\;(BL_xL_y/\Phi_0)$ is the number of vortices, and $\eta v^2/2$
is dissipation of energy per unit length of the vortex line. For a
large superconductor the efficiency, $r={\cal P}^{(tot)}_{{\rm
rad}}/{\cal P}_{{\rm dis}}$, in the case of weak disorder, is
given by $r={\Phi_0 B}/({32\pi  c\eta L_x})$. To estimate $r$ one
can use the Bardeen-Stephen formula for the drag coefficient,
$\eta = H_{c2}\Phi_0/(\rho_nc^2)$, with $\rho_n$ being the normal
state resistivity. At $L_x \sim 10\lambda$ and $B \sim 0.1
H_{c2}$, $r$ can reach $10^{-4}$. This means that at a moderate
vortex lattice velocity of 20 m/s and a cooling rate of 1 W/cm$^2$
the radiation power of 0.1 mW/cm$^2$ can be achieved. Note that
small $r$ justifies the approach in which the back effect of the
radiation on the motion of vortices is ignored.

In Conclusion, coherent electromagnetic radiation should accompany
the flux-flow state of a moving vortex lattice. The spectrum of
the radiation has discrete character and extends up to the
frequency corresponding to the superconducting gap. Thus, in
principle, this effect can be used to generate radiation in the
teraherts frequency range. Radiation power depends strongly on the
vortex lattice symmetry, velocity and degree of translational
order, providing a possible tool for characterization of moving
lattices.

We thank V. Kogan, A. Koshelev, and V. Vinokur for helpful
discussions. This work has been supported by the Department of
Energy through Contract No. W-7405-ENG-36 and Grant No.
DE-FG02-93ER45487.

\end{document}